# Lithologic Mapping of HED Terrains on Vesta using Dawn Framing Camera Color Data

Guneshwar Thangjam

Max Planck Institute for Solar System Research, Katlenburg-Lindau, Germany
Email: thangjam@mps.mpg.de

Vishnu Reddy

Max Planck Institute for Solar System Research, Katlenburg-Lindau, Germany
Department of Space Studies, University of North Dakota, Grand Forks, USA

Lucille Le Corre

Max Planck Institute for Solar System Research, Katlenburg-Lindau, Germany
Planetary Science Institute, Tucson, Arizona, USA

Andreas Nathues

Max Planck Institute for Solar System Research, Katlenburg-Lindau, Germany

Holger Sierks

Max Planck Institute for Solar System Research, Katlenburg-Lindau, Germany

Harald Hiesinger

Institut für Planetologie, Westfälische Wilhelms-Universität Münster, Munster

Jian-Yang Li

Department of Astronomy, University of Maryland, College Park, Maryland, USA

Juan A. Sanchez

Max Planck Institute for Solar System Research, Katlenburg-Lindau, Germany

Christopher T. Russell

Institute of Geophysics and Planetary Physics, University of California, Los Angeles, USA

Robert Gaskell

Planetary Science Institute, Tucson, Arizona, USA

Carol Raymond

Jet Propulsion Laboratory, California Institute of Technology, Pasadena, California, USA

**Abstract**

The surface composition of Vesta, the most massive intact basaltic object in the asteroid belt, is interesting because it provides us with an insight into magmatic differentiation of planetesimals that eventually coalesced to form the terrestrial planets. The distribution of lithologic and compositional units on the surface of Vesta provides important constraints on its petrologic evolution, impact history and its relationship with Vestoids and howardite-eucrite-diogenite (HED) meteorites. Using color parameters (band tilt and band curvature) originally developed for analyzing lunar data, we have identified and mapped HED terrains on Vesta in Dawn Framing Camera (FC) color data. The average color spectrum of Vesta is identical to that of howardite regions, suggesting an extensive mixing of surface regolith due to impact gardening over the course of solar system history. Our results confirm the hemispherical dichotomy (east-west and north-south) in albedo/color/composition that has been observed by earlier studies. The presence of diogenite-rich material in the southern hemisphere suggests that it was excavated during the formation of the Rheasilvia and Veneneia basins. Our lithologic mapping of HED regions provides direct evidence for magmatic evolution of Vesta with diogenite units in Rheasilvia forming the lower crust of a differentiated object.

# 1. Introduction

Vesta is the most massive intact basaltic object in the asteroid belt and the only surviving member of its class of protoplanets (Thomas et al. 1997; Keil 2002; McSween et al. 2011). Like the Moon, the minerals plagioclase and pyroxene, indicators of its volcanic past dominate Vesta's surface (McCord et al. 1970, 1981; Pieters 1993; Takeda 1997; Gaffey 1997). However, Vesta and the Moon have several important differences despite similar surface mineralogy. On the Moon, albedo, topography and composition are strongly related (Metzger et al. 1977; Lucey et al. 1994; Hiesinger and Head 2006). Feldspathic lunar highlands have higher albedo and are topographically higher than the lunar mare which are pyroxene-rich and have lower albedo and topography (Pieters 1993; Lucey et al. 1994, 2006). In contrast, Vesta does not show this clear correlation between albedo, composition and topography (Reddy et al. 2012b; Jaumann et al. 2012). While the topographically lower Rheasilvia basin in the southern hemisphere has higher albedo and has a predominantly diogenitic composition, several topographically higher regions also show diogenite signatures (Reddy et al. 2012b; Schenk et al. 2012). Unlike the Moon, where late stage volcanism played an important role in creating its albedo, topography and composition link (Metzger et al. 1977; Lucey et al. 1994, 2006; Hiesinger and Head 2006), Vesta's basaltic activity ceased within the first 10 million years after solar system formation (Lugmair and Shukolyukov 1998; Srinivasan et al. 1999; Coradini et al 2011; McSween et al. 2011). Eons of impacts have erased any morphological traces of its volcanic past, creating a global regolith blanket of howarditic composition (Chapman and Gaffey 1979; Bell et al. 1988; Hiroi et al. 1994; Gaffey 1997; Reddy et al. 2012b).

The NASA Dawn mission entered orbit around Vesta in July 2011, for a year-long mapping mission (Russell et al. 2011, 2012). During this period, the entire visible surface of the asteroid was imaged using a clear filter and seven color filters (Sierks et al. 2011). Recent observations (Reddy et al. 2012b; De Sanctis et al. 2012) have revealed the compositional/mineralogical diversity across Vesta, confirming several observations made from ground-based telescopes and the Hubble Space Telescope, including rotational color and albedo variations (e.g., Gaffey 1983, 1997; Dumas and Hainaut 1996; Binzel et al. 1997; Zellner et al. 2005; Carry et al. 2010; Li et al. 2010; Reddy et al. 2010). Using the Framing Camera (FC) multispectral data (0.44-0.98 µm), Reddy et al. (2012b) analyzed the color properties and albedo of Vesta. They also used 1-µm band depth and a spectral eucrite-diogenite ratio to identify regions rich in eucrites and diogenites. Using Visible and Infrared Spectrometer (VIR) hyperspectral data (0.25-5.01 µm), De Sanctis et al. (2012) analyzed regional as well as local compositional variations across the surface of Vesta. They used both 1-µm and 2-µm pyroxene band parameters, i.e., band center and band depth to interpret units in terms of diogenitic and eucritic mineralogy. These recent works from the Dawn mission show the presence of diogenites in the southern hemisphere and the eucrites in the equatorial and northen hemisphere (Reddy et al. 2012b; De Sanctis et al. 2012).

Here we present our analysis and results for identification and mapping of HED (howardite, eucrite, diogenite) terrains on Vesta using a different approach from the initial studies. We expanded our analysis by using 1-µm pyroxene band parameters that have been successfully applied to classify terrains on the Moon (Tompkins and Pieters 1999; Pieters et al. 2001; Dhingra 2007; Isaacson and Pieters 2009). Our goals are to identify and map the lithological units of the Vestan surface as constrained by HEDs, and to confirm the earlier observations, i.e., hemispherical dichotomy, albedo and band strength variations, and lithological heterogeneity, as well as an overall howarditic lithological characteristic of Vesta.

## 2. HED analyses

*2.1 1-µm Pyroxene Band Parameters and HEDs:*

Pyroxene is spectrally the most ubiquitous mineral on the surface of Vesta (e.g., McCord et al. 1970; Gaffey 1997). Eucrites are basaltic in composition, crystallized closer to the surface and contain equal amounts of low-Ca pyroxene (pigeonite) often with exsolution lamellae, and plagioclase (Mittlefehldt et al. 1998; Mayne et al. 2009; McSween et al. 2011). Diogenites are coarse-grained cumulates formed at depth and are composed mostly of orthopyroxene (Sack et al. 1991; Bowman et al. 1997; McSween et al. 2011). Howardites are regolith breccias formed by impact mixing of eucrites and diogenites (Bunch 1975; Dymek et al. 1976; Warren 1985; Warren et al. 2009). Near-IR spectra of HEDs have prominent absorption features around 1-µm and 2-µm (Gaffey 1976; Feierberg and Drake 1980; McFadden and McCord 1978; Gaffey 1997) due to the mineral pyroxene (Adams 1974, 1975; Burns 1993). The 1-µm pyroxene band center shifts to longer wavelength as the calcium content increases (Adams 1974; Cloutis and Gaffey 1991). However, it is the presence of iron ($Fe^{2+}$) in the pyroxene structure that is the cause of the two absorptions (Burns 1970, 1993; Mayne et al. 2009). Eucrites have higher calcium and iron content in their pyroxenes (54-60 mol%) than diogenites (20-33 mol%) (Mittlefehldt et al. 1998), and hence have band centers at longer wavelength than diogenites (Adams 1974; Gaffey 1976; Cloutis and Gaffey 1991).

Several authors (Tompkins and Pieters 1999; Pieters et al. 2001; Dhingra 2007; Isaacson and Pieters 2009) have successfully used pyroxene 1-µm band parameters (band curvature, band tilt and band strength) for mineralogical/compositional analyses of the lunar surface with Clementine multispectral data. These parameters can define the 1-µm absorption feature's overall shape, strength and position (Tompkins and Pieters 1999; Pieters et al. 2001; Dhingra 2007; Isaacson and Pieters 2009). Band curvature is the spectral curvature due to the 1-µm absorption feature at around 0.9 µm. Band tilt is the reflectance ratio of 0.9- and 1-µm filters while band strength is a ratio of 0.75- and 1-µm filters. Band strength that is a proxy for 1-µm band depth corresponds to the abundance of ferrous iron in a mineral (Lucey et al. 1995; Blewett et al. 1995; Tompkins and Pieters 1999; Pieters et al. 2001). Band curvature and band tilt are sensitive to mineral chemistry (Fe-rich/Ca-rich). Higher band curvature and lower band tilt are indicative of orthopyroxene, and vice versa for clinopyroxene (Pieters et al. 2001; Dhingra 2007; Isaacson and Pieters 2009). In this work, we apply these three parameters to the color spectra from the FC to constrain the surface mineralogy of Vesta and to identify terrains rich in howardites, eucrites and diogenites.

For our study we used 239 visible near-IR spectra of HED meteorites, i.e., 17 howardite, 44 eucrite, and 13 diogenite samples from the RELAB (Reflectance Experiment Laboratory) spectral database. The majority of the spectra are acquired in the wavelength range between 0.3 and 2.5 µm at 30° source angle and 0° detect angle with a spectral resolution of 5 or 10 nm. The laboratory spectra were resampled to FC filter bandpasses using the responsivity functions of the instrument (Sierks et al. 2011; Le Corre et al. 2011). Figure 1 shows the average spectra of HEDs from RELAB (1A) along with their resampled FC color spectra (1B), and 1-µm band parameters (1C) that will be used for this study. Band parameters were extracted from RELAB spectra using the following equations from Isaacson and Pieters (2009) that have been adapted for Dawn FC data:

Band Curvature (BC) = $(R_{0.749}+R_{0.964})/R_{0.918}$           Eq. 1

Band Tilt (BT) = $R_{0.918}/R_{0.964}$           Eq. 2

Band Strength (BS) = $R_{0.964}/R_{0.749}$           Eq. 3

R is the reflectance value at the corresponding wavelengths.

*2.2 Band Curvature (BC):*

We applied the band curvature equation (Eq. 1) to RELAB spectra of HEDs in an effort to differentiate the three meteorite types. According to our analyses, eucrites have a band curvature range of 2.18-3.4 with an average of 2.75±0.22, while diogenites are in the range of 2.78-4.44 (80% in the range of 3.4-4.5) with an average of 3.39±0.32. Howardites have intermediate values (2.48-3.33) with an average value of 2.90±0.22. BCs for the samples with ≤25 μm particle size (howardites-13, eucrites-27, diogenites-11 samples) show a similar trend (92% of eucrites in the range of 2.6-3.1 and 63% of diogenites in the range of 3.4-3.8). Based on these analyses, eucrites have lower BCs, while diogenites have higher values, with intermediate values for howardites. The higher band curvature in diogenites is due to 1-μm absorption features located at shorter wavelength due to lower total iron/calcium abundance while lower band curvature in eucrites is due to higher iron/calcium abundance that shifts the 1-μm absorption feature to longer wavelength.

  To study the effect of particle size on BC, we analyzed color spectra of three samples ALHA76005 (eucrite), EETA79002 (diogenite) and EET87503 (howardite) in different particle size ranges (≤25, 25-45, 45-75, 75-125, 125-250, 250-500 μm). Figure 2 (A, B, C) shows the spectra normalized to unity at 0.75 μm for the three samples and figure 3 (A, B, C) shows the observations for the effects of grain-size for the three samples. We found that the band curvature of HEDs is affected by particle size. Depending on the particle size range, BCs varies between 11-19% with reference to the minimum value in their size ranges. The samples in the 25-45 and 45-75 μm particle size range are most affected while ≤25 μm size range is least affected (Fig. 3). Despite particle size effects, BC is still a reliable indicator for iron abundance and can be used for differentiating eucrites and diogenites as shown in Figure 4A.

*2.3 Band Tilt (BT):*

Applying the band tilt equation (Eq. 2) to our samples, we found that the band tilt of eucrites ranges between 0.90-1.1 with an average of 0.96±0.02. The band tilts of diogenites range between 0.73-0.89 (95% in the range of 0.81-0.89) with an average of 0.84±0.03. Howardites have intermediate values (0.83-0.97) with an average value of 0.92±0.03. For the samples with particle sizes ≤25 μm, 81% of eucrites are in the range of 0.93-0.99 while 81% of diogenites are in the range of 0.82-0.87. Based on our analysis, eucrites have stronger BTs while diogenites have lower values and howardites have intermediate values. Higher band tilts in eucrites are due to 1-μm absorption features located at longer wavelengths due to higher iron/calcium abundances while lower band tilts in diogenites are due to lower iron/calcium abundances shifting the 1-μm absorption feature to shorter wavelength. Higher band tilt is observed in olivines too (Pieters et al. 2001; Isaacson and Pieters 2009), but olivine is rare in HED meteorites (Sack et al. 1991; Fowler et al. 1994; Mittlefehldt 1994; McSween et al. 2011) and not yet reported on Vesta. Therefore higher band tilt is referred to eucrites for the present analyses.

Unlike band curvature, band tilt is less affected by particle size and varies only 2-7%. The samples with the largest particle size-range (250-500 µm) show the widest variations and the effect decreases as the particle size decreases (Fig. 3). Because of the negligible particle size effect, band tilt can be used as a very effective parameter for differentiating eucrites from diogenites.

*2.4 Band Strength (BS):*

Applying band strength equation (Eq. 3), howardites, eucrites and diogenites are observed to have wide ranges of BS values. Eucrites are in the range of 0.44-0.84 with an average of 0.62±0.08, while diogenites are between 0.35-0.68 with an average of 0.49±0.07. Howardites have intermediate values (0.50-0.61) with an average value of 0.61±0.06. For the samples with ≤25 µm particle size-range, similar distributions are observed. Despite overlapping BS ranges, diogenites have overall lower BS values, implying deeper band depths compared to eucrites and howardites.

Depending on the particle size, BS varies between 17% and 23%. In the observed range, samples with 25-45 and 45-75 µm particle size-ranges have the deepest band depth and the depths decrease as the particle sizes increase or decrease (Fig. 2, 3). Therefore, band depth is not uniquely affected by composition, but also by particle size. The band depth on Vesta is also affected by phase angle, temperature, iron abundance and the presence of opaques (Hiroi et al. 1994; Duffard et al. 2005; Reddy et al. 2012a, 2012b, 2012c).

*2.5 Combined approach of band curvature (BC) and band tilt (BT):*

Based on our analysis of spectra of HED meteorites, we determined that band curvature and band tilt are the most robust spectral parameters to distinguish eucrites from diogenites. A scatter plot of BC vs. BT with an approximate outline of HED-regions shows distinct spatial distributions for eucrites and diogenites with howardites falling in between (Fig. 4A, B). Irrespective of particle size, 19% and 52% of howardites fall in the diogenite-region and the eucrite-region, respectively (Fig. 4A). A similar trend is observed when we use spectra of samples with a particle size-range of ≤25 µm (Fig. 4B). Here 25% and 68% of howardites are in the diogenite region and eucrite-region, respectively. This suggests that howardites, in general, are more eucrite-rich (50-58%) than diogenite-rich (19-25%). This result is consistent with observed compositional variations in Vestoids (Binzel and Xu 1993; Burbine et al. 2009; Reddy et al. 2011).

*2.6 Band Strength (BS) and Albedo at 0.749-µm:*

Despite variations in albedo at 0.75-µm and band depth of HEDs, a correlation between band depth and albedo exists (Reddy et al. 2012a, 2012b, 2012c). The scatter plot of BS and albedo for the HEDs irrespective of the particle size (Fig. 5A) shows the variability of band depth as well as albedo with an overall deeper band depth for the diogenites. However, the scatter plot for the samples considered in the particle size-range ≤25 µm shows a linear correlation with an r-factor (correlation coefficient) of 0.66 (Fig. 5B). We also analyzed the effect of particle size on albedo for the three samples with different grain size-ranges and found that albedo decreases with increasing particle size, i.e., from ≤25 μm (54-70%) to a minimum for the 250-500 μm grain size-range. In contrast, band depth does not directly correlate with particle size, but the maximum band depth is observed between 25-45 and 45-75 μm, and decreases as the particle size increases or decreases. The smallest particle size range (≤25 μm) has the weakest band depth but highest albedo variation (54-70%) of the three samples. Given the lack of clear correlation between albedo and band strength with composition and the observed effect of opaques in the form of dark material on band depth on Vesta (Reddy et al. 2012c; McCord et al. 2012), we are not using these observations for the present study.

## 3. Data Reduction and Analysis

The Dawn FC acquired multispectral images using a clear filter and seven band-pass filters in the wavelength range of 0.43-μm to 0.96-μm with high spatial resolution up to 16 m/pixel (Sierks et al. 2011; Reddy et al. 2012b). Basic calibration (bias, dark, and flats) is accomplished prior to the generation of higher-level data products in the Integrated Software for Imagers and Spectrometers (ISIS) developed and maintained by the USGS (United States Geological Survey) (Anderson et al. 2004). The data are photometrically corrected using Hapke functions derived from disk integrated, ground-based telescopic and Dawn spacecraft observations of Vesta and Vestoids (Li et al. 2012). The data are map-projected and co-registered to align the seven color frames to create the color cubes. A detailed discussion of the FC data processing pipeline is presented in the supplementary section of Reddy et al. (2012b). For our study, data acquired during the approach phase (more specifically during the Rotational Characterization 3b, i.e., RC3b) of the mission were used with a pixel scale of ~480 m/pixel. All maps shown here are in the Claudia coordinate system (Russell et al. 2011, 2012) in simple cylindrical projection and span the latitudinal range from 75°S to 26°N.

We used the image analysis software ENVI 4.8 to analyze the global maps. The three band parameters equations (band curvature, band tilt and band strength) are calculated with the band math application and are color-coded. Apart from the three band parameter maps (Fig. 6B-D), an albedo map (the reflectance image at 0.75-μm) was created to provide the context for the other maps (Fig. 6A). To analyze and map the HED regions on Vesta, 2D scatter plot of band curvature vs. band tilt and the band parameters' global images as well as the trend of HEDs spectral observations were used. Using the Regions of Interests (ROIs) tool in ENVI 4.8 software, HED regions identified in the previous step were exported separately for further analysis.

## 4. FC data analyses

The global albedo mosaic at 0.75-μm (Fig. 6A) shows wide variations across the surface of Vesta. The geometric albedo varies between 0.10 and 0.67 (Reddy et al. 2012b). Most of the regions in the eastern hemisphere and few regions in the western hemisphere have lower albedo while regions in the southern hemisphere and few in the northern hemisphere have higher albedos. Such albedo variations on hemispheric scale are consistent with the north-south and east-west dichotomies (Bobrovnikoff 1929; Haupt 1958; Gehrels 1967; Dumas and Hainaut 1996; Gaffey 1997; Binzel et al. 1997; Rivkin et al., 2006; Li et al. 2010; Reddy et al. 2010, 2012b).

*4.1 Band Curvature (BC):*

The band curvature global mosaic from the RC3b phase is shown in Figure 6B. Band curvature across the surface of Vesta has a range of 1.61-3.96, with an average value 2.59±0.09. Regions with the weak BCs (blue) have lower albedos and are largely confined to the eastern hemisphere and a few limited areas in the western hemisphere. In contrast, the southern hemisphere and a few regions in the northern hemisphere (probably recent impact craters and their ejecta) have higher values of BC (red/yellow/green). These regions have higher albedos, suggesting a link between BC and albedo. As observed from the analysis of HED meteorites, higher values of BC are indicative of diogenites, while lower values are typical of eucrites. Our BC map of Vesta suggests that the southern hemisphere and few limited regions in the northern hemisphere are dominated by diogenite-rich material, while equatorial regions and the northern hemisphere are dominated by eucrite-rich materials. The intermediate values of BC are indicative of howardites, which are the result of collisional mixing of eucrite-rich and diogenite-rich regions. However, since BC is also affected by particle size besides the composition, it is worthy of caution in interpreting compositions.

*4.2 Band Tilt (BT):*

Figure 6C shows the BT map of Vesta from the RC3b phase of the mission. The color-coding and values of BT are inverted (Fig. 6C) to be consistent with the BC map (Fig. 6B). Band tilt measured on the surface of Vesta has a range between 0.71-1.69, with an average value of 1.04±0.01. The band tilt map shows the hemispherical dichotomy with the lower albedo and lower BT regions in the eastern hemisphere, while the higher albedo regions in the southern hemisphere have higher BT values. As observed from the analysis of HED meteorites, BT is a very robust parameter to distinguish eucrites from diogenite with negligible effects of particle size differences. Therefore we conclude that the equatorial and the northern hemisphere regions are dominated by eucrite-rich material, while the regions in the southern hemisphere are more diogenitic.

*4.3 Band Strength (BS):*

The band strength map of Vesta is shown in Figure 6D. Similar to the band tilt map (Fig. 6C), the BS color-code and values are inverted to show areas of deeper 1-μm band depth, in red. Band strength ranges between 0.56-2.76 with an average value of 1.49±0.07. The BS map shows variations in the strength of the 1-μm pyroxene absorption band across the surface of Vesta. The southern hemisphere has higher albedos and deeper band strengths than the northern hemisphere, which has lower albedos and weaker band strengths. Similarly, the low albedo regions in the eastern hemisphere and a few regions in the western hemisphere also show weaker band strengths. This suggests a correlation between band strength and albedo. The presence of carbonaceous chondrite materials in the northern hemisphere, largely in the western hemisphere causes weaker band depths and lower albedos (Reddy et al. 2012b, 2012c; McCord et al. 2012). As observed from the analysis of HED meteorites, diogenites typically show stronger BSs than eucrites. However, particle size and the presence of opaques and metal have significant impact on this parameter (Hiroi et al. 1994; Duffard et al. 2005; Reddy et al. 2012b, 2012c; McCord et al. 2012). Our BS analysis coupled with BT and BC data suggests that diogenite is the dominant material in the southern hemisphere of Vesta (Li et al. 2010; De Sanctis et al. 2012; Reddy et al. 2010, 2012b).

*4.4 Lithological mapping using pyroxene band parameters:*

We qualitatively analyzed the Vestan surface using the pyroxene band parameters (Fig. 6B-D) in an effort to identify howardite-, eucrite- and diogenite-rich regions. The lithological units are mapped over the global albedo mosaic (Fig. 6E) considering the observations of the ratio images (preferentially band tilt, and then band curvature) as well as the observations of the HED meteorites. The mapped HED units are also marked in the scatter plot of band curvature vs. band tilt (Fig. 4C). Although the band parameter values observed from HED meteorites and FC data are slightly mismatched, the trends observed in HED meteorites (Fig. 4 A, B) and FC global mosaic (Fig. 4C) are generally similar. An exact comparison of values from the laboratory and actual observations may not be possible because of the uncertainties and the difference primarily in the spectra of laboratory samples and actual surface materials (e.g., Mustard et al. 1993; Tompkins and Pieters 1999). Considering the uncertainty limits of FC filters spectral responses and their calibration in the RC3b data (filters with the center wavelengths 0.43-, 0.65-, 0.75-, 0.96-µm could be affected upto 2% and filters with the center wavelengths 0.83-, 0.92-µm up to 4%), deviations from the RC3b values are calculated to find the maximum probable error for the band parameters (Fig. 4C). The mismatch in the values remains to be explained in the near future with the availability of better higher resolution FC data with entire coverage of Vesta. As seen from the scatter plot and the lithological map, the howardite-rich regions fall between eucrite-rich and diogenite-rich regions. Like in the HED data, we also observed a larger overlap between the eucrite- and howardite-rich units compared to the diogenite- and howardite-rich regions.

We compared color spectra of the HED regions and the average color spectrum of Vesta from the RC3b data. Figure 7 shows the average spectra for each lithology normalized to unity at 0.75 µm. These spectra show variations in 1-µm band parameters due to composition. The average color spectrum of Vesta is consistent with howardite-rich regions, confirming the howarditic nature of the Vestan surface (Chapman and Gaffey 1979; Bell et al. 1988; Hiroi et al. 1994; Gaffey 1997; Zellner et al. 2005; Delaney 2009; Carry et al. 2010; De Sanctis et al. 2012). The average band tilt for eucrite-, diogenite- and howardite-rich regions is 0.97±0.02, 0.93±0.02, and 0.96±0.01, respectively while the band tilt value for the entire surface is 0.96±0.01. Similarly, the average value of band curvature is 2.5±0.15, 2.79±0.07, 2.6±0.18 for eucrite-, diogenite-, and howardite-rich regions and 2.6±0.2 for the entire surface. Thus BT and BC values for surface average are consistent with howardite-rich regions, confirming the earlier spectral match (Fig. 7). As derived from the band parameters, diogenite-rich regions are generally in the southern hemisphere while eucrite- and howardite-rich regions are in the equatorial and northern hemisphere (Li et al. 2010; De Sanctis et al. 2012; Reddy et al. 2010, 2012b). Again, the diogenite-rich regions are generally characterized by relatively deeper band depth and higher albedo as compared to eucrite-rich and howardite-rich regions (Li et al. 2010; De Sanctis et al. 2012; Reddy et al. 2010, 2012b). Thus, the hemispherical dichotomy in albedo and composition is also observed in the band parameters (Bobrovnikoff 1929; Haupt 1958; Gehrels 1967; Dumas and Hainaut 1996; Gaffey 1997; Binzel et al. 1997; Rivkin et al. 2006; Li et al. 2010; Reddy et al. 2010, 2012b). The presence of diogenite-rich materials dominantly in the southern hemisphere suggests that it was excavated during the formation of the giant south pole basins, particularly the Rheasilvia basin (Reddy et al. 2012b; McSween et al. 2013).

## 5. Conclusions

Our analysis of Dawn FC color data using 1-µm pyroxene band parameters has confirmed several findings from previous works as discussed above and has provided new insight into the distribution of eucrite- and diogenite-rich material on the surface of Vesta. Our study reveals the following:

- We have successfully applied lunar band parameter analysis technique to Dawn FC data to identify terrains rich in eucrites and diogenites.
- HED laboratory spectra show distinct variations in pyroxene absorption spectral parameters (band strength, band tilt and band curvature) that can be used to interpret surface mineralogy.
- Band tilt and band curvature are strongly influenced by pyroxene chemistry (Fe-rich/Ca-rich) and are robust indicators to identify eucrite- or diogenite-rich regions.
- Band strength is strongly influenced by particle size and abundance of opaques (carbonaceous chondrite materials) on the surface and is a less reliable indicator for the abundance of ferrous iron.
- Vesta shows hemispherical dichotomies (north-south and east-west) in albedo and color that are strongly related to the excavation of diogenite-rich lower mantle material during the formation of the Rheasilvia and Veneneia basins.
- The average color spectrum of Vesta is similar to the average spectrum of howardite regions and the band tilt and band curvature parameters are also identical to those of howardites.

**Acknowledgment**

We thank the Dawn team for the development, cruise, orbital insertion, and operations of the Dawn spacecraft at Vesta. The Framing Camera project is financially supported by the Max Planck Society and the German Space Agency, DLR. We also thank the Dawn at Vesta Participating Scientist Program for funding the research. A portion of this work was performed at the Jet Propulsion Laboratory, California Institute of Technology, under contract with NASA. Dawn data is archived with the NASA Planetary Data System. This study uses 239 HED meteorites spectra from the RELAB spectral database at Brown University and we acknowledged the concerned PIs and the RELAB team for their effort. We thank Thomas Burbine, the anonymous reviewer and Harry Y. McSween for their helpful and constructive comments.

**Figure Caption:**

Figure 1. (**A**) Average spectra of HED meteorites obtained from RELAB spectral database in the visible/near-infrared wavelength range showing 1-μm pyroxene absorption feature. (**B**) Average HED spectra resampled to Dawn Framing Camera (FC) wavelength range using FC bandpasses and filter responsivity. (**C**) Average spectra of the HEDs normalized to unity at 0.75-μm showing the three 1-μm band parameters, i.e., band curvature, band tilt and band strength.

Figure 2. Normalized spectra of a (**A**) howardite (EET87503), (**B**) eucrite (ALHA76005, and (**C**) diogenite (EETA79002) showing the effect of particle size. Spectra normalized to unity at 0.75 μm and samples with particle size ranges of ≤25 μm, 25 to 45 μm, 45 to 75 μm, 75 to 125 μm, 125 to 250 μm, and 250 to 500 μm are shown.

Figure 3. Spectral parameters and albedo at 0.749-μm of (**A**) howardite (EET87503), (**B**) ALHA76005 (eucrite), (C) EETA79002 (diogenite) showing the effect of particle size on the band parameters and albedo.

Figure 4. (**A**) Scatter plot of band curvature vs. band tilt derived from the resampled spectra of the HEDs. 41 howardites, 157 eucrites, and 41 diogenites of all particle sizes are shown in this plot. (**B**) Samples with particle size from ≤25 μm (13 howardites, 27 eucrites, 11 diogenites) are shown in this plot. (**C**) Scatter plot using the RC3b global mosaic shows the eucrite-rich and diogenite-rich regions, with the howardite-rich regions falling in between.

Figure 5. (**A**) Scatter plot of band strength (BS) vs. albedo at 0.749-μm derived from the resampled spectra of the HEDs (41 howardites, 157 eucrites, and 41 diogenites) irrespective of particle size. (**B**) Samples with particle size ≤25 μm (13 howardites, 27 eucrites, 11 diogenites).

Figure 6. Global mosaic of Vesta acquired during the approach phase (RC3b) at a resolution of 480 m/pixel in the Claudia coordinate system with a simple cylindrical projection. (A) Albedo image at 0.75-μm shows variations in albedo across the surface and 1-μm band parameter maps i.e., (**B**) band curvature (BC), (**C**) band tilt (BT) and (**D**) band strength (BS) with rainbow-color code and their values. The original values of BT and BS and the color ramp are inverted to show diogenite-rich regions as reddish/yellowish and eucrite-rich/howardite-rich regions as bluish/greenish color in the band parameter's image. (**E**) Lithological units mapped as eucrite-rich, diogenite-rich and howardite-rich. Yellow lines (dashed and dotted) are approximate outlines for the south pole basins (the Rheasilvia and the Veneneia).

Figure 7. Average color spectra of howardite-rich, eucrite-rich and diogenite-rich regions from Fig. 6E normalized to unity at 0.749-μm. The average color spectrum of the howardite-rich region and the one for the whole Vestan surface are overlapped and similar in nature.

**Figures:**

Fig. 1: Lithological Mapping of Vesta

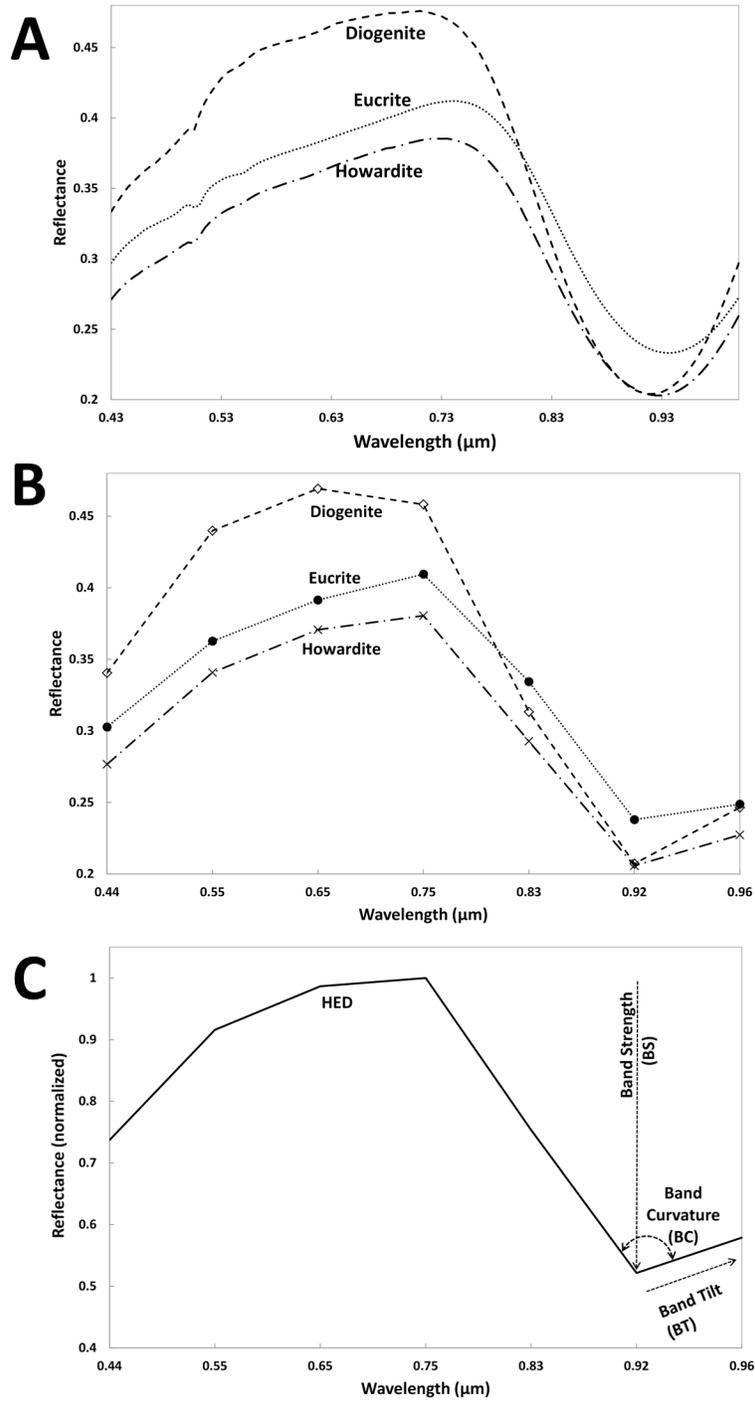

Fig. 2: Lithological Mapping of Vesta

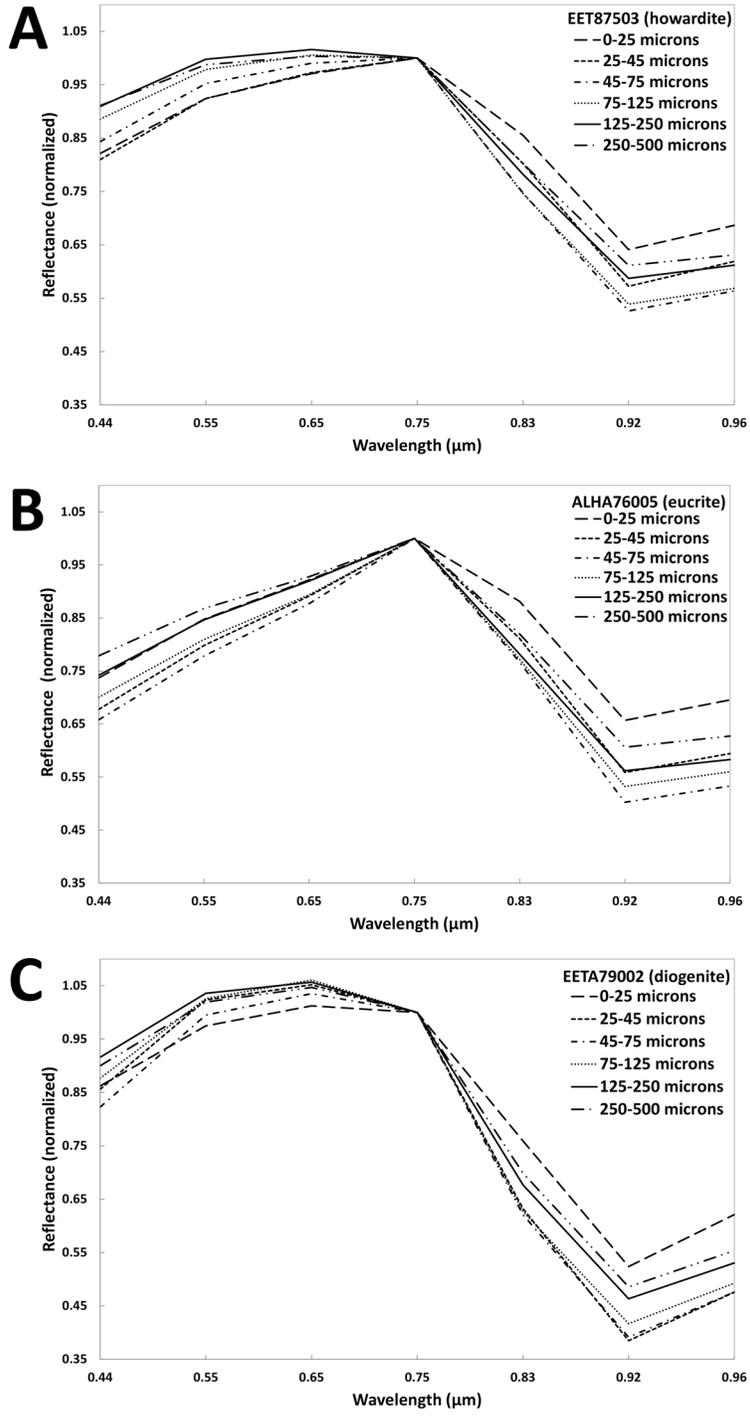

Fig. 3: Lithological Mapping of Vesta

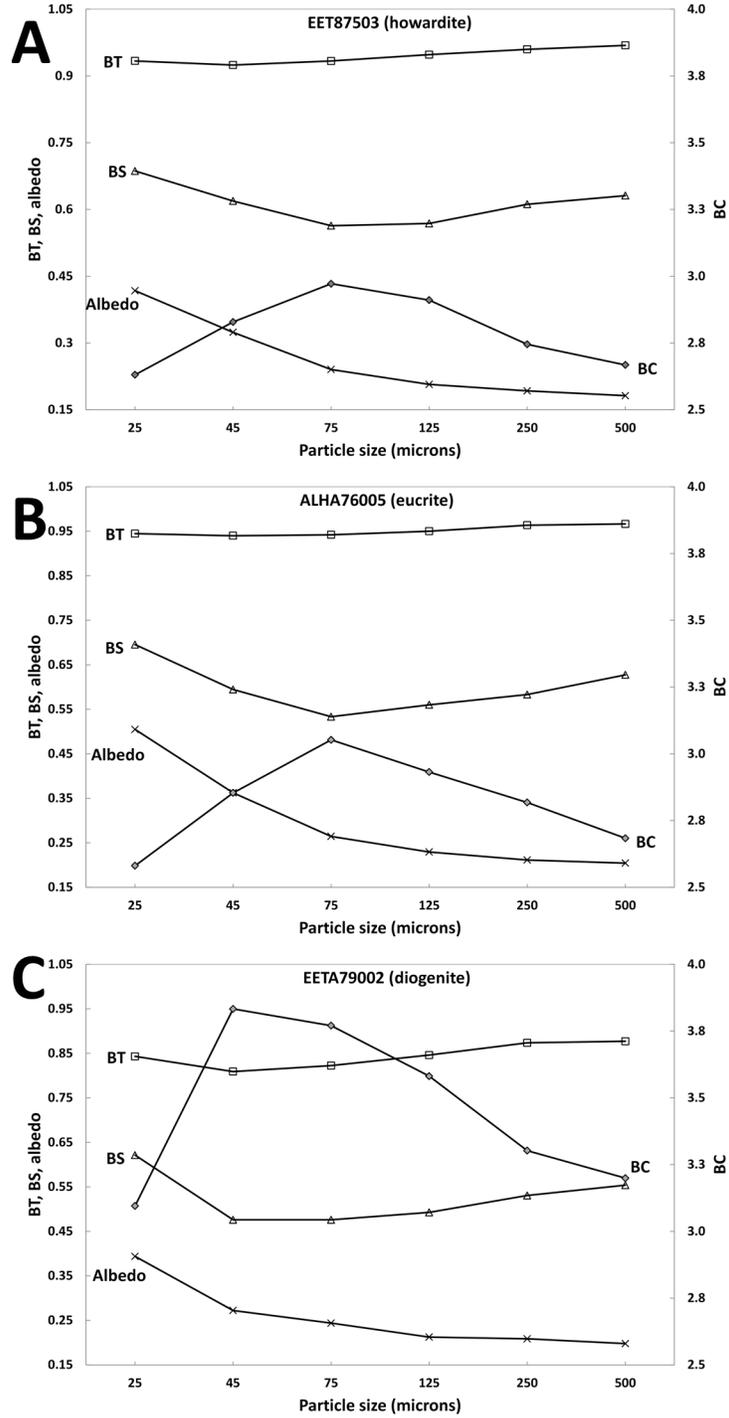

Fig. 4: Lithological Mapping of Vesta

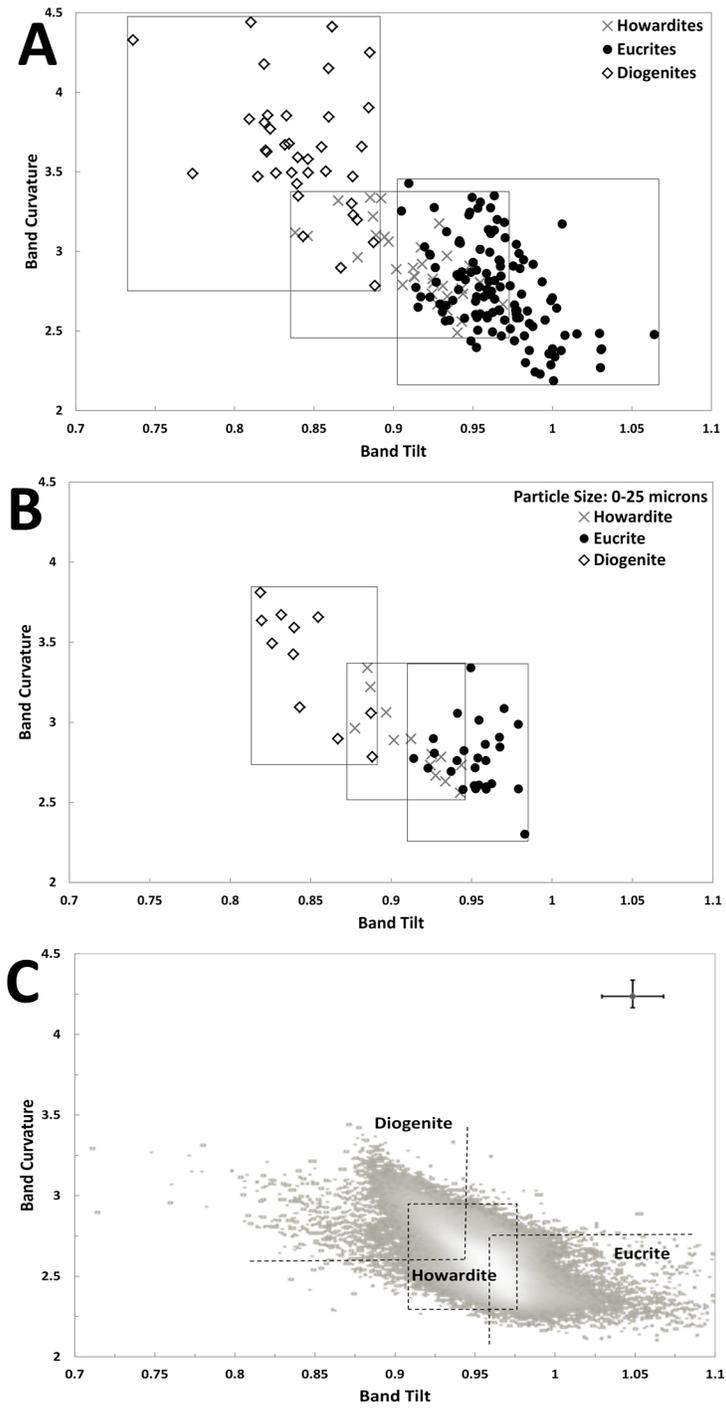

Fig. 5: Lithological Mapping of Vesta

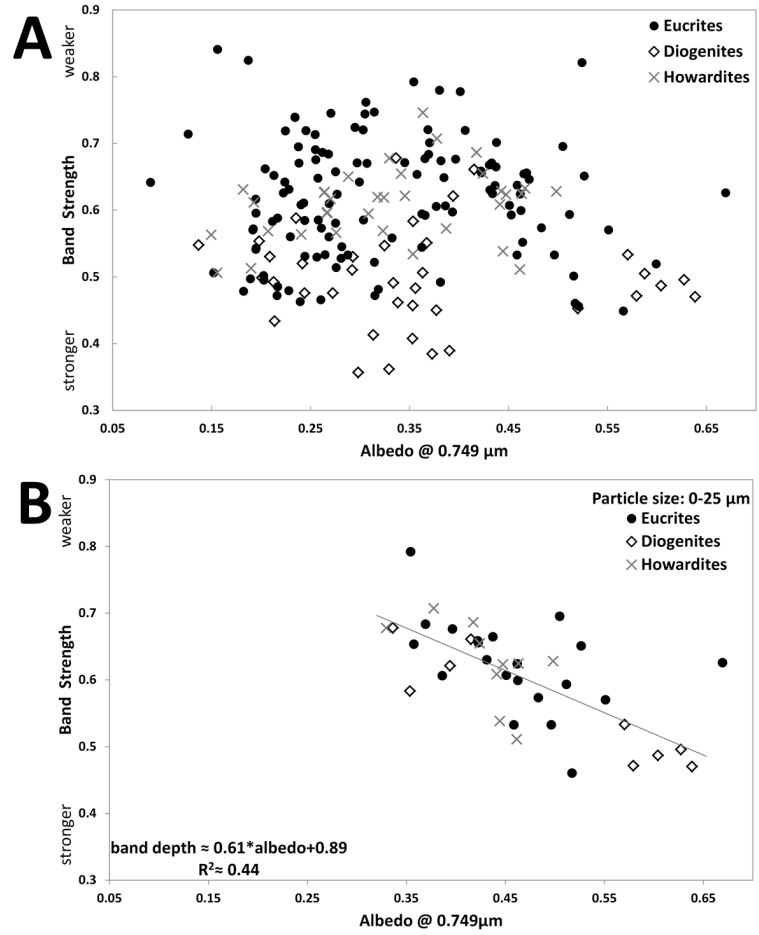

Fig. 6: Lithological Mapping of Vesta

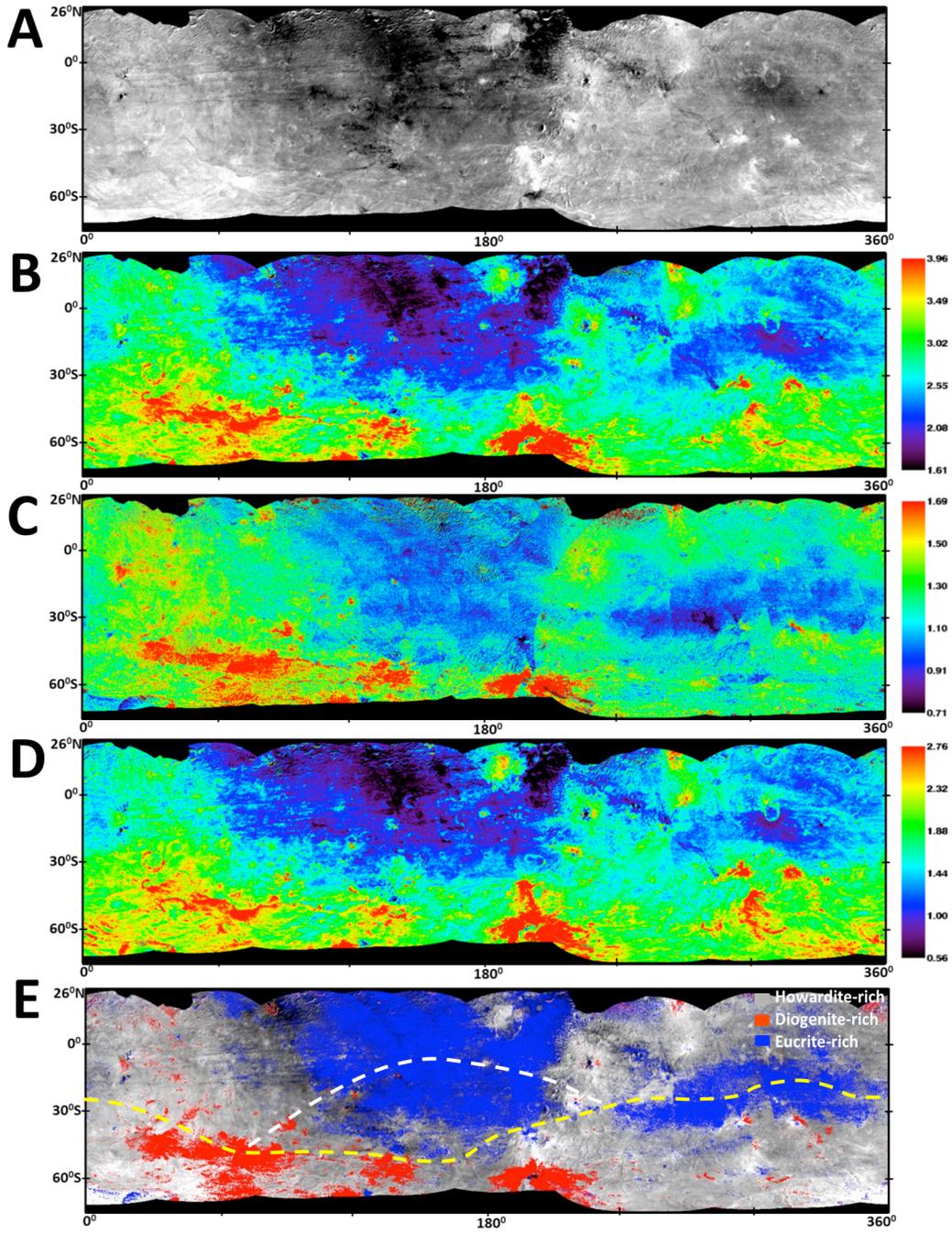

Fig. 7: Lithological Mapping of Vesta

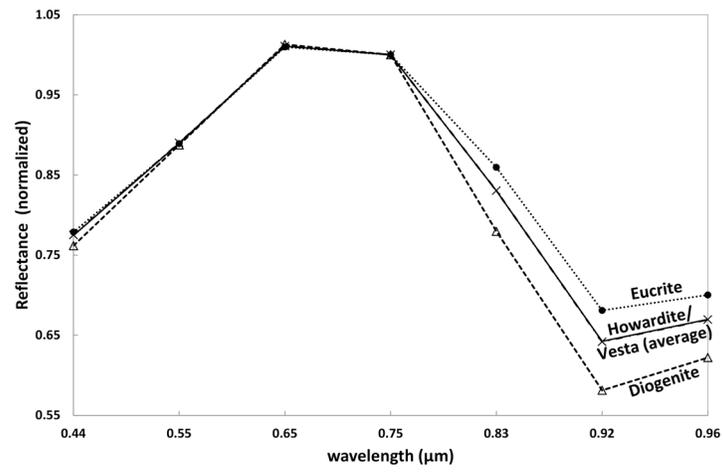